\documentstyle[12pt]{article}
\begin{document}

\topmargin 0pt
\oddsidemargin=-0.4truecm
\evensidemargin=-0.4truecm

\vspace*{-2.0cm}

\vspace*{0.5cm}
\newcommand{\smU}{{\scriptscriptstyle U}}
\newcommand{\sm}[1]{{\scriptscriptstyle #1}}
\def\d{\partial}
\def\l{\left(}
\def\r{\right)}
\newcommand{\e}{\mathop{\rm e}\nolimits}
\newcommand{\Tr}{\mathop{\rm Tr}\nolimits}
\renewcommand{\Im}{\mathop{\rm Im}\nolimits}
\renewcommand{\Re}{\mathop{\rm Re}\nolimits}
\newcommand{\be}{\begin{equation}}
\newcommand{\ee}{\end{equation}}
\newcommand{\ba}{\begin{eqnarray}}
\newcommand{\ea}{\end{eqnarray}}


\begin{center}
  {\Large\bf Quantum Cosmology}\footnote{Lecture at
   NATO ASI ``Structure Formation in the Universe'',
   Cambridge, 26 July -- 6 August, 1999}  \\
  \medskip V.~A.~Rubakov,
\footnote{E-mail:    \verb|rubakov@ms2.inr.ac.ru|}\\
  \medskip
  {\small
     Institute for Nuclear Research of
         the Russian Academy of Sciences,\\  60th October Anniversary
  Prospect, 7a, Moscow, 117312, 
Russia\\
 and\\
           University of Cambridge,\\
           Isaac Newton Institute for Mathematical Sciences,\\
           20 Clarkson Road, Cambridge, CB3 0EH, U.K. 

  }
  
\end{center}

\begin{abstract}
We comment on two issues in quantum cosmology, in the context
of the Wheeler--De Witt equation and wave function of the Universe:
(i) arrow of time and interpretation of the wave function in the
classically allowed regions; (ii) stability of an approximation of
the Born--Oppenheimer type in classically forbidden regions of 
the scale factor.
\end{abstract}

\section{Introduction}

Most of the evolution of the Universe is likely to have
proceeded classically, in the sense that the dominant phenomenon 
is the classical expansion while quantum fluctuations are small 
and can be treated as perturbations. Prior to this stage, however,
genuinely quantum phenomena almost certainly took place.
Although it is not clear whether they have left any observable
footprints, it is of interest to try to understand them, as
this may shed light on such issues as initial conditions for
classical cosmology (are inflationary initial data natural?
how long was the inflationary epoch? is open Universe consistent
with inflation?), properties of space-time near the
cosmological singularity, origin of coupling constants, etc.

To describe the Universe at its quantum phase, one ultimately has to
deal with full quantum gravity theory, well beyond the
Einstein gravity. It is, however, legitimate to take more modest
attitude and consider quantum phenomena below the Planck 
(or string) energy scale. Then the quantized Einstein gravity
(plus quantized matter {\it fields}) provides an 
effective ``low energy'' description which must be tractable, 
at least in principle, within quantum field theory framework.
Processes that should be possible to consider in this way are not
necessarily perturbative, as the example of tunneling in
quantum mechanics and field theory shows.

Surprisingly, not so many phenomena are well understood even
within this modest approach. Perhaps the most clear process is the
decay of a metastable vacuum \cite{ColDL}. It has been clarified 
recently \cite{RSib} that at least in the range of parameters where
the treatment of space-time 
in terms of background de Sitter metrics
is reliable, the Coleman--De Luccia instanton indeed describes 
the false vacuum decay,
provided the quantum fluctuations above the classical false vacuum
are in de Sitter-invariant (conformal) vacuum state. This is in
full accord with the results of Refs. \cite{ColDL,GuthWein}.
It is likely that this conclusion holds also when the quantum 
properties of metrics are taken into account.
Furthermore,  the Hawking--Moss instanton \cite{HMoss}
can be interpreted \cite{RSib}
as a limiting case of constrained instantons
that describe the false vacuum decay in an
appropriate region of parameter space,
again in agreement with previous analyses \cite{StarCDL,GonLin}.
Hence, there emerges a coherent picture of the false vacuum decay
with gravity effects included.

One may try to apply laws of quantum mechanics to the Universe as a
whole, and consider the wave function of the Universe. Although
research in this direction began more than 30 years ago \cite{WDW},
the situation here is  still intriguing and controversial.
The main purpose of this contribution is to make a few comments
on this subject. Namely, we will discuss which analogies to
ordinary quantum mechanics are likely to work in quantum cosmology, 
and which are rather misleading. We begin with quantum mechanics,
and only then turn to the wave function of the Universe.

\section{Wave function in quantum mechanics} 
To set the stage, let us consider a quantum mechanical system
with two dynamical coordinates, $x$ and $y$. Let the Hamiltonian be
\[
    \hat{H} = \hat{H}_0 + \hat{H}_y
\]
where
\[
    \hat{H}_0 = \frac{1}{2} \hat{p}_x^2 + V_0(x)
\]
\[
    \hat{H}_y = \frac{1}{2} \hat{p}_y^2 
                + \frac{1}{2} \omega^2(x) y^2
                +\frac{1}{4} \lambda(x) y^4 + \dots
\]
Let us assume that the potential $V_0 (x)$ is such that the 
motion along the coordinate $x$
is semiclassical, while the dynamics along the coordinate $y$ 
can be treated in perturbation theory about the semiclassical motion
along $x$. This approach is close in spirit to the Born--Oppenheimer
approximation. We will consider solutions to the {\it stationary}
Schr\"odinger equation with fixed energy $E$.

Let us first discuss the dynamics in the classically
allowed region of $x$, where $E>V_0(x)$. In this region,
there are two sets of solutions with the semiclassical parts 
of the wave functions equal to
\begin{equation}
   \Psi \propto \mbox{e}^{+iS(x)}
\label{7*}
\end{equation}
and
\[
     \Psi \propto \mbox{e}^{-iS(x)}
\]
where
\[
    S(x) = \int^x~dx'~\sqrt{2(E-V_0(x'))}
\]
These two sets of solutions correspond to motion right and left,
respectively. Note that this 
interpretation is based on the fact that there exists
{\it extrinsic time} $t$ inherent in the problem: 
the complete, time-dependent wave functions 
are $\exp (-iEt + iS(x))$ and $\exp (-iEt - iS(x))$;
the wave packets constructed out of the wave functions of these
two types indeed move right and left, respectively, as
$t$ increases.

Let us now consider the dynamics along the coordinate $y$,
still using the time-independent Schr\"odinger equation 
in the allowed region of $x$. This is done for, say, right-moving
system by writing, instead of eq.(\ref{7*}),
\[
    \Psi(x,y) = \frac{1}{\sqrt{p_x(x)}} \tilde{\Psi}(x,y)
                \mbox{e}^{iS(x)}
\]
where $p_x = \partial S/ \partial x$. To the first order in
$\hbar$ one obtains that the time-independent Schr\"odinger
equation reduces to
\begin{equation}
     i \frac{\partial \tilde{\Psi}}{\partial x}
       \frac{\partial S}{\partial x} = \hat{H}_y \tilde{\Psi}  
\label{8+}
\end{equation}
This can be cast into the form of {\it time-dependent}
Schr\"odinger equation by changing variables from $x$ to
$\tau$ related by $x=x_c(\tau)$, where 
 $x_c(\tau)$ is the solution of the classical 
equation of motion for $x$ in ``time'' $\tau$,
which has energy $E$ and obeys
\begin{equation}
     \frac{\partial S}{\partial x}(x=x_c)= 
            \frac{\partial x_c}{\partial \tau}
\label{8*}
\end{equation}
After this change 
of variables, $\tilde{\Psi}$ becomes a function of $y$ and $\tau$
and obeys the following equation,
\begin{equation}
   i \frac{\partial \tilde{\Psi}(y;\tau)}{\partial \tau} =
   \hat{H}_y(\hat{y},\hat{p}_y; \tau) \tilde{\Psi}(y;\tau)
\label{9*}
\end{equation}
where the explicit dependence of $\hat{H}_y$ on $\tau$ comes from 
$x_c(\tau)$. We see that there have emerged intrinsic time $\tau$
which parameterizes the classical trajectory $x_c(\tau)$ and also
the $y$-dependent part of the wave function. We note again that 
in quantum mechanics, the
arrow of intrinsic time, which is set by the sign convention in 
eq.(\ref{8*}), is determined by the arrow of extrinsic time $t$.

Note also that one is free to choose any 
representation for operators $\hat{y}$ and $\hat{p}_y$ and write,
instead of eq.(\ref{9*}),
\begin{equation}
  i\frac{\partial |\tilde{\Psi}\rangle}{\partial{\tau}}
    = \hat{H}_y(\tau)  |\tilde{\Psi}\rangle
\label{9**}
\end{equation}
To solve this equation, one may find convenient
to switch to the
Heisenberg representation, as usual.

We now turn to the discussion of the 
region of $x$ where the classical motion is forbidden and the
system has to tunnel. To simplify formulas, we set $E=0$ in what
follows. If the system tunnels from left to right, the dominant
semiclassical wave function is
\[
    \Psi \propto \mbox{e}^{-S(x)}
\]
where $S(x) = \int^x~dx'~\sqrt{2V_0(x')}$ and obeys the 
following equation,
\[
    -\frac{1}{2}\left(\frac{\partial S}{\partial x}\right)^2
     + V_0(x) = 0
\]
This equation may be formally considered as the classical 
Hamilton--Jacobi equation in Euclidean (``imaginary'') time.
The zero energy classical trajectory $x_c(\tau)$
in Euclidean time $\tau$ obeys
\[
    \frac{ d^2 x_c}{d \tau^2} 
      = +\frac{\partial V_0}{\partial x}(x=x_c)
\]
and hence
\[ 
   \frac{dx_c}{d\tau} = \frac{\partial S}{\partial x}(x=x_c)
\]
Then $S(x)$ can be calculated as the value of
the Euclidean action along this trajectory.

To find the equation governing the dynamics along  $y$-direction
in the classically forbidden region of $x$, we again write
\[
    \Psi(x,y) = \frac{1}{\sqrt{p_x}} \tilde{\Psi}(x,y)
                \mbox{e}^{-S(x)}
\]
and obtain, changing variables from $x$ to $\tau$,  $x=x_c(\tau)$,
that $\tilde{\Psi}$ obeys the time-dependent Schr\"odinger
equation, now in Euclidean time,
\begin{equation}
   \frac{\partial \tilde{\Psi}(y;\tau)}{\partial \tau} =
    - \hat{H}_y(\hat{y}, \hat{p}_y; \tau) \tilde{\Psi} (y;\tau)
\label{11*}
\end{equation}
The minus sign on the right hand side of this equation 
{\it is crucial} for the stability of the approximation we use.
Indeed, the system described by eq.(\ref{11*}) tends to de-excite, rather 
than excite, as ``time'' $\tau$ increases, so that the part
 $\tilde{\Psi}$ of the wave function remains always subdominant
as compared to the leading semiclassical exponential. The physics 
behind this property is quite clear: we consider tunneling at 
fixed energy, so the de-excitation of fluctuations along $y$ means
the transfer of energy to the tunneling subsystem, which makes 
tunneling (exponentially) more probable. Inversely, if fluctuations 
along $y$ get excited, the kinetic energy along $x$ decreases,
and tunneling gets suppressed stronger.

\section{Wave function of the Universe}

To discuss specific aspects of quantum cosmology, let us consider 
the closed Friedmann--Robertson--Walker Universe with the scale 
factor $a$. Let us introduce the cosmological constant $\Lambda$,
minimal scalar field $\phi (x)$ with a scalar potential $V(\phi)$
and also massless conformal scalar field. We are going to 
treat the dynamics of the scale factor in a semiclassical manner;
in this respect $a$ is analogous to the variable $x$ of the previous
section. The minimal scalar field (as well as gravitons) will be
considered within perturbation theory, so each of the modes  
$\phi_{\bf k}$ will be analogous to the variable $y$ of the
previous section.

The basic equation in quantum cosmology is the Wheeler--De Witt 
equation, which in our case reads
\begin{equation}
    \left[ -\frac{1}{2} \hat{p}_a^2 - \frac{1}{2} a^2
     + \Lambda a^4 + \hat{H}_{\phi} \right] \Psi
    = -\epsilon \Psi
\label{14*}
\end{equation}
where we have set $3M_{Pl}^2/16\pi = 1$ and 
ignored the operator ordering
problems which are irrelevant for our discussion.
Here
\[
   \hat{H}_{\phi} = 
     \int~\frac{d^3x}{2\pi^2}~\left[ \frac{1}{2a^2}\hat{p}_{\phi}^2
     +\frac{a^2}{2} (\partial_i \hat{\phi})^2 +
     a^4 V(\hat{\phi})\right]
\]
is the term due to the minimal scalar field; at the classical 
level $\hat{H}_{\phi}$ is the energy of matter defined with respect to
conformal time. The non-negative constant $\epsilon$ on the right 
hand side of eq.(\ref{14*}) is the contribution of the conformal 
scalar field; the only purpose of introducing the latter field is to allow
for non-zero $\epsilon$. We do not consider gravitons in what follows,
as they are similar to the quanta of the minimal scalar field $\phi$.

In the spirit of the Born--Oppenheimer approximation, let us first 
neglect the conformal energy of the field $\phi$, i.e., omit the term
$\hat{H}_{\phi}$ in eq.(\ref{14*}). Then the Wheeler--De Witt equation
takes the form of the time-independent Schr\"odinger equation in quantum
mechanics of one generalized coordinate $a$ with energy
$\epsilon$ and potential
\[
   U(a) = \frac{1}{2} a^2 - \Lambda a^4
\]
At $16\Lambda^2 \epsilon < 1$, there are two classically allowed regions:
at small $a$ ($0<a^2<[1-\sqrt{1-16\Lambda^2\epsilon}]/4\Lambda$)
and at large $a$
($\infty>a^2>[1+\sqrt{1-16\Lambda^2\epsilon}]/4\Lambda$). 
At the classical level, the former region corresponds to an
expanding and recollapsing Friedmann-like closed Universe, while the
latter corresponds to the de Sitter-like behavior. As $\epsilon \to 0$, 
the first classically allowed region disappears, while the second becomes
exactly de Sitter.

In between these two regions, classical evolution is impossible
(if one neglects $\hat{H}_{\phi}$),  and one has to consider classically
forbidden ``motion''. Let us discuss classically allowed and classically
forbidden regions separately.

\subsection{Classically allowed region:  issue of arrow of time}

To be specific, let us consider classically allowed de Sitter-like 
region where the scale factor $a$ is large. In the leading order, there are
again two types of semiclassical wave functions,
\begin{equation}
       \Psi \propto \mbox{e}^{-iS(a)}
\label{16*}
\end{equation}
and
\begin{equation}
       \Psi \propto \mbox{e}^{+iS(a)}
\label{16**}
\end{equation}
where 
\[
    S(a) = \int^a~da'~\sqrt{2(\epsilon - U(a'))}
\]
Classically, the momentum is related to the derivative of the conformal 
factor with respect to conformal time,
\[
   \frac{da}{d\eta}=-p_a
\]
For the two semiclassical wave functions one has
\[
   \hat{p}_a \Psi = 
     \left(\mp \frac{\partial S}{\partial a}\right) \Psi
\]
where upper and 
lower signs refer to eq.(\ref{16*}) and eq.(\ref{16**}),
respectively. Hence, one is tempted to interpret the wave functions
(\ref{16*}) and (\ref{16**}) as describing expanding and 
contracting Universes, respectively.  Indeed, the Hartle--Hawking
wave function \cite{HH} that in the allowed region is a superposition
\begin{equation}
    \Psi_{HH} \propto \mbox{e}^{-iS(a)} + \mbox{e}^{+iS(a)}
\label{17*}
\end{equation}
is often interpreted as describing a collapsing and re-expanding 
de Sitter-like Universe. 
Similar interpretation is often given to the Linde wave function
\cite{Tun1}.
On the other hand, the tunneling wave functions
\cite{Tun2,Tun3,VRold} which contain one wave in the allowed region,
\[
   \Psi_{tun} \propto  \mbox{e}^{-iS(a)}
\]
are often assumed to be the only ones that
correspond to an expanding, but not contracting, Universe; this is,
at least partially, the basis for the
tunneling interpretation.

An important difference with conventional quantum mechanics
is, however, the absence of extrinsic time in quantum cosmology.
Hence, the arrow of intrinsic time has yet to be determined.
In other words, there is no {\it a priori} 
reason to interpret the wave 
functions (\ref{16*}) and (\ref{16**}) as describing expanding and
contracting Universes, respectively.
The sign of the semiclassical exponent does not by itself determine
the arrow of time.

Were the scale factor the only dynamical variable,
it would be impossible to decide whether, say, the wave function
(\ref{16*}) corresponds to expanding or contracting Universe.
If the matter fields (and/or gravitons) are included, 
this should be possible. Before discussing this point, let us 
derive the equation for the wave function describing 
matter \cite{LR,HaHa,Banks}, again in the spirit of the 
Born--Oppenheimer approximation.

Let us extend the wave functions (\ref{16*}) and (\ref{16**}) to contain
the dependence on the matter variables,
\begin{equation}
   | \Psi (a) \rangle = \frac{1}{\sqrt{p_a}} \mbox{e}^{\mp iS(a)}
            | \tilde{\Psi} (a) \rangle
\label{19*}
\end{equation}
where at given $a$ both $| \Psi (a) \rangle$ and  
$|\tilde{\Psi} (a) \rangle$ belong to the Hilbert space in which
$\hat{\phi}({\bf x})$ and $\hat{p}_{\phi}({\bf x})$ act. As an example,
one may (but does not have to) 
choose the generalized coordinate representation;
then  $| \Psi (a) \rangle$ becomes a function $\Psi(\{\phi_{\bf k}\};a)$
of the Fourier components of $\phi$.  In the first order in $\hbar$ 
one obtains from eq.(\ref{14*})
 \begin{equation}
   \pm i \sqrt{\epsilon - U(a)}
   \frac{\partial |\tilde{\Psi} (a) \rangle}{\partial a}
    = \hat{H}_{\phi} |\tilde{\Psi} (a) \rangle
\label{20*}
\end{equation}
in complete analogy to eq.(\ref{8+}). 

The arrow of time is determined 
now by where (at what $a$) and which initial conditions are
imposed on $|\tilde{\Psi} (a) \rangle$. As an example, let us assume 
that the initial conditions 
for the evolution in real intrinsic time
are imposed at small $a$ (at the turning
point
$a^2=[1+\sqrt{1-16\Lambda^2\epsilon}]/4\Lambda$), and that
at that point $|\tilde{\Psi} \rangle$ describes smooth distribution of
the scalar field. This type of initial data are characteristic, in
particular, to the Hartle--Hawking no-boundary wave function.
As $a$ increases, the system will become more and more disordered,
independently of the sign in eq.(\ref{20*}). With thermodynamical
arrow of time, {\it both} wave functions (\ref{19*}) will describe 
expanding Universe.

If, with these initial conditions, 
one changes variables from $a$ to $\eta$ using
\[
   \frac{\partial a}{\partial \eta} = \sqrt{\epsilon - U(a)}
\]
then $\eta$ increases with $a$, so that $\eta$ is the conformal
intrinsic time, independently of the choice of sign in eq.(\ref{19*}).
In the case of positive sign, eq.(\ref{20*}) becomes the conventional
Schr\"odinger equation for quantized matter in the
expanding Universe,
\begin{equation}
   i \frac{\partial |\tilde{\Psi}  \rangle}{\partial \eta}
    = \hat{H}_{\phi}(\eta) |\tilde{\Psi} \rangle    
\label{21*}
\end{equation}
where the matter Hamiltonian depends on $\eta$ through $a(\eta)$.
On the other hand, in the case of negative  sign one obtains
``wrong sign'' Schr\"odinger equation,
\[
   i \frac{\partial |\tilde{\Psi}  \rangle}{\partial \eta}
    = -\hat{H}_{\phi}(\eta) |\tilde{\Psi} \rangle    
\]
This little problem is easily cured by considering, instead of 
$|\tilde{\Psi}  \rangle$, its $T$-conjugate, 
$|\tilde{\Psi}^{(T)}  \rangle$; if the generalized coordinate 
representation is chosen for $|\tilde{\Psi}  \rangle$, then $T$-conjugation
is merely complex conjugation, 
$\tilde{\Psi}^{(T)}(\phi_{\bf k};\eta) 
=\tilde{\Psi}^{*}(\phi_{\bf k};\eta)$.
The $T$-conjugate wave function obeys conventional Schr\"odinger 
equation, but with $CP$-transformed Hamiltonian. Hence, the 
interpretation 
of {\it both} wave functions (\ref{19*}) as describing
the expanding Universe is self-consistent; the only peculiarity is that
the wave function $\mbox{e}^{+iS}|\tilde{\Psi} \rangle$ corresponds to
the Universe in which matter is $CP$-conjugate.
In particular, we
argue that both components of the Hartle--Hawking wave function
(\ref{17*}) correspond to expanding Universes.

In more generic cases (in particular, 
when the matter degrees of freedom 
cannot be  treated perturbatively, see, e.g., Refs. \cite{Kam,Un}
and references therein), the situation may be much more
complicated. Still, the arrow of time is generally expected
to be one of the 
key issues in the interpretation of the wave function of the Universe.

\subsection{Classically forbidden region: issue of
stability of the Born-Oppenheimer approximation}

We now consider the region of the scale factor that is 
classically forbidden in the absense of $\hat{H}_{\phi}$,
i.e., $a_1<a<a_2$, where
\[
   a_{1,2}^2= \frac{1 \mp \sqrt{1-16\Lambda^2\epsilon}}{4\Lambda}
\]
If $\hat{H}_{\phi}$ is switched off, there are two semiclassical
solutions to the Wheeler--De Witt equation,
\begin{equation}
     \Psi \propto \mbox{e}^{-S(a)}
\label{23*}
\end{equation}
and
\begin{equation}
     \Psi \propto \mbox{e}^{+S(a)}
\label{23**}
\end{equation}
where
\[
   S(a) = \int_{a_1}^a~da'~\sqrt{2(U(a')-\epsilon)}
\]
is defined in such a way that it always increases at large $a$.
The wave function (\ref{23*}) decays as $a$ increases, so it may
be interpreted as describing tunneling from classically allowed
Friedmann region to de Sitter-like one. It is convenient to 
introduce Euclidean conformal time  parameter $\tau$ and consider
Euclidean trajectory $a_c(\tau)$ obeying
\[
    \frac{da_c}{d\tau} = \frac{\partial S}{\partial a} (a=a_c)
\]
At $\epsilon=0$ the Euclidean four-geometry corresponding to
this trajectory is a four-sphere, the standard de Sitter instanton.

Let us now turn on the scalar field Hamiltonian $\hat{H}_{\phi}$, and
try to apply the procedure of the Born--Oppenheimer type.
We write, instead of eq.(\ref{23*}), for the wave function 
decaying at large $a$,
\[
  | \Psi (a) \rangle = \frac{1}{\sqrt{p_a}} \mbox{e}^{-S(a)}
            | \tilde{\Psi} (a) \rangle   
\]
and obtain in the first order in $\hbar$ that 
$ | \tilde{\Psi} (a) \rangle $ obeys the ``wrong sign'' 
Euclidean Schr\"odinger equation \cite{VRold} 
\begin{equation}
  \frac{\partial |\tilde{\Psi}(\tau)  \rangle}{\partial \tau}
    = + \hat{H}_{\phi}(\tau) |\tilde{\Psi}(\tau) \rangle 
\label{24*}
\end{equation}
where the change of variables from $a$ to $\tau$ with
$a=a_c(\tau)$ has been performed.  The sign on the right
hand side of eq.(\ref{24*}) is opposite to that appearing in the
usual quantum mechanics, eq.(\ref{11*}), and is directly
related to the sign of $\hat{p}_a^2$-term in the Wheeler--De Witt 
equation (\ref{14*}).

The ``wrong'' sign in eq.(\ref{24*}) implies that the approximation
we use is in fact unstable, if generic ``initial'' conditions are 
imposed at small $a$, say, at $a=a_1$. Note that imposing 
initial conditions in this way is natural if one interprets the 
wave function decaying at large $a$ as describing tunneling from small
to large $a$. The formal reason for the instability of the
approximation is that the degrees of freedom of the scalar field get
excited as $a$ increases in the forbidden region. The rate at which this
excitation occurs is generically high \cite{VRold}, and the approximation
breaks down well before $a$ gets close to the second turning point $a_2$.

In the path integral framework, breaking of the Born--Oppenheimer-type
approximation for the wave function decaying at large $a$ is also
manifest \cite{TurokHbr}. This wave function corresponds to the
Euclidean path integral with ``wrong'' sign of the action,
\[
   \int~Dg~D\phi~\mbox{e}^{+S[g,\phi]}  
\]
The instanton action then gives the factor $\mbox{e}^{-S_{inst}}$, 
but the integral over $\phi$ (and gravitons) diverges.

The physics behind this instability is that tunneling of a Universe
filled with matter is exponentially
more probable as compared to empty Universe. Hence, the matter 
degrees of freedom tend to get excited 
in the forbidden region, thus making
tunneling easier. Note that this property is 
peculiar to quantum cosmology: in quantum mechanics the 
situation is opposite, as we discussed in the previous section.

There are exceptional cases in which matter degrees of freedom do not
get excited in the forbidden region, e.g., because of symmetry. 
In our model this would be the case if $\epsilon=0$
and the scalar field $\phi$ was in the
de Sitter-invariant state,
cf. Ref. \cite{VachVil}. Such cases do not seem generic, however.

Breaking of the Born--Oppenheimer approximation does not necessarily
mean that tunneling-like transitions 
from small $a$ to large $a$  with generic state of matter at 
small $a$ do not make sense. Rather, it is the semiclassical expansion 
that does not work in this case, so the state of the Universe after
the transition may be quite 
unusual.
Presently, neither the properties of this state, nor the properties
of the wave function in the forbidden region are understood
(except for special cases mentioned above).

The situation is different for the wave functions increasing 
towards large $a$, eq.(\ref{23**}). In that case the matter wave 
function obeys the usual Euclidean Schr\"odinger equation,
$ \partial |\tilde{\Psi}(\tau)  \rangle / \partial \tau
    = - \hat{H}_{\phi}(\tau) |\tilde{\Psi}(\tau) \rangle $,
where $\tau$ is still assumed to increase with $a$. Hence, it
is possible to impose fairly general initial conditions at small $a$,
and the approximation will not break down. In particular, the 
Hartle--Hawking wave function is a legitimate approximate solution
to the Wheeler--De Witt equation in the forbidden region.
This is in accord with the path-integral treatment: the increasing
wave function (\ref{23**}) corresponds to the standard sign of
the Euclidean action in the path integral.

The non-semiclassical behavior of the tunneling wave functions,
signalled by the instability of the Born--Oppenheimer-type
approximation, is a special, and potentially interesting,
feature of quantum cosmology. It is a challenging technical problem
to develop techniques adequate to this situation. It is not excluded 
also that the properties of the tunneling wave functions are rich
and complex, and that  
understanding them may shed light on the
beginning of our Universe.

The author is indebted to A. Albrecht, J. Goldstone, N. Turok,
W. Unruh and A. Vilenkin for helpful discussions.

%
%
%

\end{document}